\documentclass[12pt]{article}
\usepackage{amsmath,amssymb}
\usepackage{epsfig}
\usepackage{graphicx}
\usepackage{a4}
\usepackage{latexsym}
\usepackage{cite}

\graphicspath{{Pics/}}
\DeclareGraphicsExtensions{.eps,.ps}

\usepackage{pslatex}
\usepackage{txfonts}
\usepackage[latin1]{inputenc}
\usepackage[T1]{fontenc}

\newcommand{\lsim}{\raisebox{-0.07cm}{$\:\:\stackrel{<}{{\scriptstyle \sim}}\:\: $} }

\def\ca{{C^{}_A}}
\def\cas{{C^{\,2}_A}}
\def\cf{{C^{}_F}}
\def\cfs{{C^{\,2}_F}}
\def\nf{{n^{}_{\! f}}}
\def\b0{\beta_0}

\def\m2{m^2}
\def\Q2{Q^2}
\def\mus{{\mu^{\,2}}}
\def\muf{{\mu^{}_f}}
\def\mufs{{\mu^{\,2}_f}}
\def\mur{{\mu^{}_r}}
\def\murs{{\mu^{\,2}_r}}
\def\alphas{{\alpha_s}}

\def\rs{{r_s}}
\def\Lbeta{L_\beta}
\def\deltas4{{\delta(s_4)}}
\def\D#1{{D_{#1}}}
\def\z#1{{\zeta_{#1}}}
\def\Li#1{{\rm Li}_{#1}}
\def\t1{{t_{1}}}
\def\u1{{u_{1}}}

\begin{document}
\setlength{\parskip}{0.2cm}
\setlength{\baselineskip}{0.535cm}

\begin{titlepage}
\noindent
DESY 08-163\\
SFB/CPP-08-91 \\
November 2008 \\
\vspace{1.2cm}
\begin{center}
\Large {\bf 
  Higher order QCD corrections to charged-lepton deep-inelastic
  scattering and global fits of parton distributions 
}

\vspace{2.4cm}
\large
S. Alekhin$^{\, a,}$\footnote{{\bf e-mail}: sergey.alekhin@ihep.ru} 
and S. Moch$^{\, b,}$\footnote{{\bf e-mail}: sven-olaf.moch@desy.de} \\
\vspace{1.0cm}
\normalsize
{\it $^a$Institute for High Energy Physics \\
\vspace{0.1cm}
142281 Protvino, Moscow region, Russia}\\
\vspace{0.5cm}
{\it $^b$Deutsches Elektronensynchrotron DESY \\
\vspace{0.1cm}
Platanenallee 6, D--15738 Zeuthen, Germany}\\
\vspace{0.5cm}
\vfill
\large {\bf Abstract}
\vspace{-0.2cm}
\end{center}
We study the perturbative QCD corrections to heavy-quark structure functions of charged-lepton deep-inelastic scattering 
and their impact on global fits of parton distributions.
We include the logarithmically enhanced terms near threshold due to soft
gluon resummation in the QCD corrections at next-to-next-to-leading order.
We demonstrate that this approximation is sufficient to describe the available HERA data 
in most parts of the kinematic region.
The threshold-enhanced next-to-next-to-leading order corrections improve the agreement 
between predictions based on global fits of the parton distribution functions 
and the HERA collider data even in the small-$x$ region. 
\vspace{1.5cm}
\end{titlepage}

\noindent
In this letter we study structure functions in deep-inelastic scattering (DIS) of charged leptons off nucleons.
We focus on the production of heavy quarks, like e.g., charm, 
which proceeds within perturbative Quantum Chromodynamics (QCD) predominantly through 
boson-gluon fusion~\cite{Witten:1975bh,Gluck:1979aw}.
This is a reaction of great interest because at moderate momentum transfer $Q$, 
it provides a direct probe of the gluon content of the nucleon over a wide range of Bjorken $x$.
In the present work, we specifically include higher order QCD corrections to 
DIS heavy-quark production and we wish to determine their impact on our information 
about the parton distribution functions (PDFs) and, especially, the gluon PDF.

Our motivation stems from the high statistics data for the charm structure function $F_2^c$  
provided by the HERA experiments~\cite{Aktas:2005iw,Chekanov:2003rb}, 
where $F_2^c$ accounts for a large fraction (up to $30\%$) of the total structure function $F_2$. 
The presently available DIS data allows for high precision extractions of PDFs 
in global fits~\cite{Alekhin:2002fv,Alekhin:2006zm}. 
The treatment of the charm contribution in these fits is an important issue as it can induce
potentially large effects also in the PDFs of light quarks and the gluon
obtained from these global fits (see e.g. the recent review~\cite{Thorne:2008xf}).

In the standard factorization approach the heavy-quark contribution to the DIS
structure functions can be written as a convolution of PDFs and coefficient functions, 
\begin{equation}
  \label{eq:totalF2c}
  F_k(x,\Q2,\m2) =
  {\alphas\, e_q^2\, \Q2 \over 4 \pi^2 \m2} \,\,
  \sum\limits_{i = q,{\bar{q}},g} \,\,
  \int\limits_{a x}^{1}\,
  dz \,\, f_{i}\left(z, \mufs\right)\,\,
  c_{i, k}\left(\eta(x/z),\xi,\mufs,\murs\right)
  \, ,
\end{equation}
where $a = 1 + 4 \m2/\Q2$ and $m$, $e_q$ are the heavy-quark mass and charge.
The strong coupling constant at the renormalization scale $\mur$ is denoted $\alphas = \alphas(\mur)$ 
and the PDFs for the parton of flavor $i$ are $f_{i}(x,\mufs)$ at the factorization scale $\muf$.
Both scales $\muf$ and $\mur$ are assumed equal throughout this work, i.e. $\mu = \muf = \mur$.
The kinematical variables $\eta$ and $\xi$ in Eq.~(\ref{eq:totalF2c}) are given as 
\begin{equation}
  \label{eq:eta-xi-def}
  \eta = {s \over 4 \m2} - 1\, ,
  \qquad\qquad\qquad
  \xi = {\Q2 \over \m2}\, ,
\end{equation}
and the partonic center-of-mass energy $s=\Q2 (z/x-1)$ in Eq.~(\ref{eq:totalF2c}).
The coefficient functions of the hard partonic scattering process enjoy an expansion in $\alphas$ as 
\begin{equation}
  \label{eq:coeff-exp}
  c_{i, k}(\eta,\xi,\mus) \,=\,
  \sum\limits_{k=0}^{\infty}\, (4 \pi \alphas(\mu))^k \,
  \sum\limits_{l=0}^{k} c^{(k,l)}_{i,k}(\eta, \xi) \ln^l\frac{\mus}{\m2}
  \, .
\end{equation}

The perturbative QCD predictions for the coefficient functions to the leading order (LO) are long known~\cite{Witten:1975bh,Gluck:1979aw}.
The next-to-leading order (NLO) radiative corrections are available since more than 15 years~\cite{Laenen:1992zk}. 
Likewise, the massless coefficient functions for the light-quark content of the DIS structure functions 
have also been calculated to the next-to-next-to-leading order (NNLO) 
some time ago~\cite{Kazakov:1992xj,vanNeerven:1991nn,Zijlstra:1991qc,Zijlstra:1992kj,Moch:1999eb}.
More recently, the scale evolution of the PDFs has been matching the NNLO accuracy~\cite{Moch:2004pa,Vogt:2004mw}.

Although the full heavy-quark coefficient functions at two loops are the only unknown 
for a complete NNLO analysis, much can already be said about these terms at present.
Because the structure functions for massive quarks contain two hard scales, that is 
the momentum transfer $Q$ and the heavy-quark mass $m$, the study of particular
kinematical limits yields valuable information.

For instance, at asymptotic values $\Q2, \mus \gg \m2$ one may treat the heavy quark as effectively massless.
As an upshot, large logarithms $\ln(\mus/\m2)$ are summed over in the PDF evolution~\cite{Shifman:1977yb}.
Accordingly, the higher order coefficient functions~\cite{Laenen:1992zk} in Eq.~(\ref{eq:coeff-exp}) 
take asymptotic forms~\cite{Buza:1995ie,Bierenbaum:2007qe} 
and, together with the PDFs, require matching. 
This is the standard procedure when changing the description 
from QCD with $\nf$ light flavors and a heavy quark to a theory with $\nf + 1$ light quarks. 
Thus, a so-called variable flavor number scheme (VFNS) 
has to describe this transition in the effective number of light flavors 
and, moreover, a general-mass formalism for a VFNS has to be consistent with QCD
factorization, see Refs.~\cite{Chuvakin:1999nx,Tung:2001mv}.
However, the VFNS formalism cannot be routinely extrapolated to the region of $Q\sim m$, 
where the power corrections due to the heavy-quark mass effects break the factorization 
and therefore the massive quarks cannot be considered as partonic constituents of the nucleon. 
In this kinematical region, one strictly applies QCD with $\nf$ light flavors and one heavy quark, 
thus working in the so-called fixed flavor number scheme (FFNS).
The VFNS can be used for the whole kinematics of the existing DIS data  
only if it is matched in order to provide a smooth transition to the FFNS at small $Q$. 
Details of this transition cannot be derived within the VFNS framework and are therefore subject to model assumptions 
(see Ref.~\cite{Thorne:2008xf} for the review of the modern state of art and history of this modeling). 
On the other hand, in the FFNS the large logarithms $\ln(\mus/\m2)$ are contained in the higher order corrections 
to the coefficient functions. 
Thus, the basic motivation for the use of a VFNS weakens, at least for the realistic DIS kinematics~\cite{Gluck:1993dpa}.

Near threshold, for $s \simeq 4 \m2$ or equivalently $\eta \ll 1$, higher order perturbative corrections are much enhanced.
The coefficient functions in Eq.~(\ref{eq:coeff-exp}) exhibit large
double logarithms $\alphas^l \ln^{2l}\beta$, with $\beta=\sqrt{1-4\m2/s}$ being the velocity of the heavy quark 
and these Sudakov logarithms can be resummed to all orders in perturbation theory.
Currently this has been achieved to the next-to-leading logarithmic (NLL) accuracy
and the resummed result can be employed to generate approximate results
at NNLO in QCD (see e.g.~\cite{Laenen:1998kp}).
In the present work, we focus on the impact of these approximate NNLO
corrections due to soft gluons and assess their impact 
on global fits of PDFs in the FFNS.
To that end, it is instructive to express the convolution in Eq.~(\ref{eq:totalF2c}) as an 
integration over the partonic variable $\eta$. 
In this way, we obtain with $z(\eta)=x(1+4(\eta+1)/\xi)$  
\begin{equation}
  \label{eq:F2c-eta}
  F_k(x,\Q2,\m2) = 
  {\alphas\, e_q^2 \over \pi^2} \,\,
  \sum\limits_{i = q,{\bar{q}},g} \,\,
  \int\limits_{0}^{\eta_{max}}\,
  d\eta \,\, x\, f_{i}\left(z(\eta), \mus\right)\,\,
  c_{i, k}\left(\eta,\xi,\mus\right)
  \, ,
\end{equation}
where the integration is bounded by $\eta_{max}=\xi/4(1/x-1)-1$.

\begin{figure}[htb]
  \begin{center}
    \includegraphics[width=14.0cm]{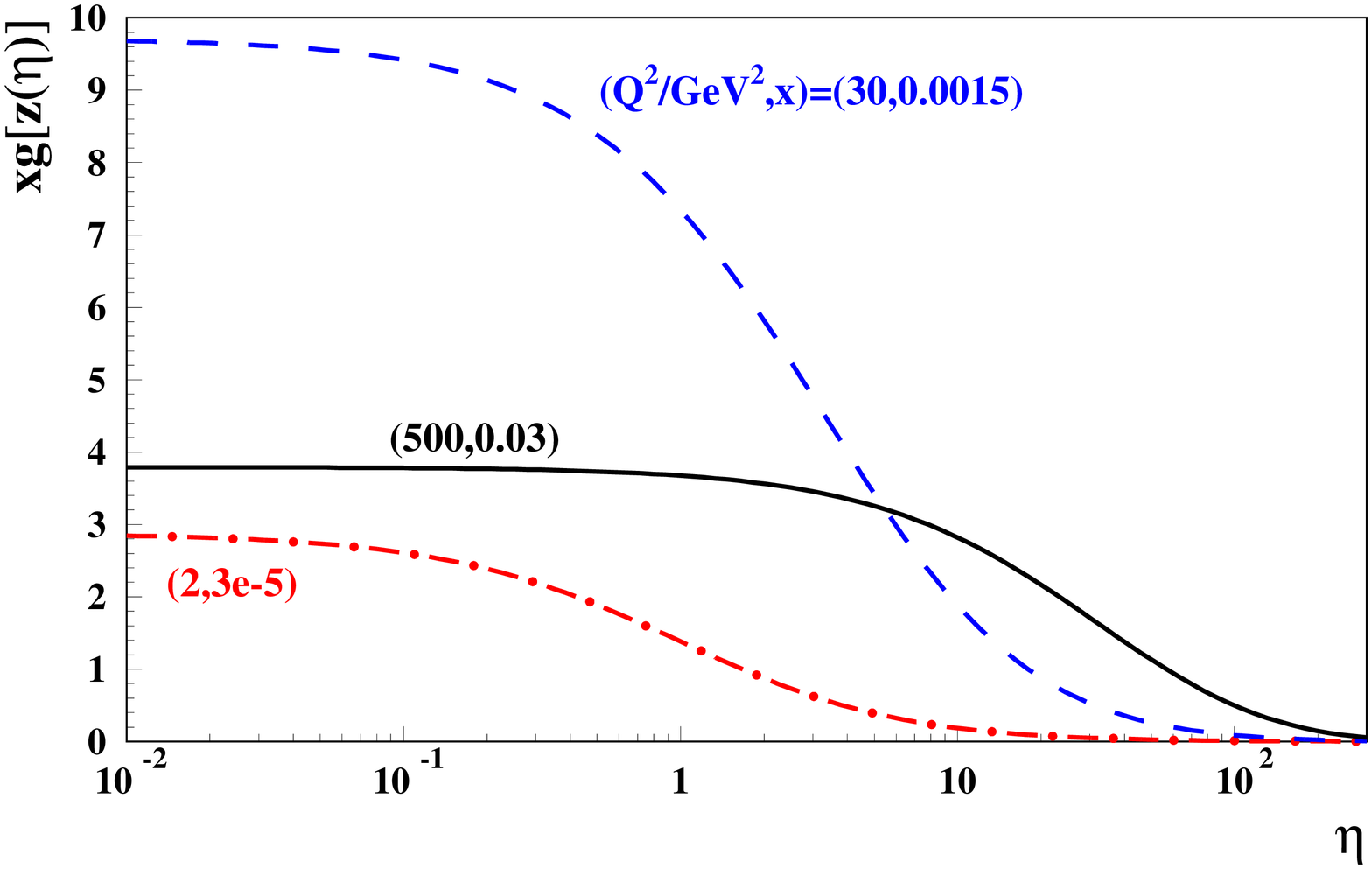}
    \vspace*{-5mm}
    \caption{ \small
      The $\eta$-dependence of the 
      gluon distribution $xg(z)$ for representative kinematics of the HERA 
      collider experiments (see Eq.~(\ref{eq:F2c-eta})).
      \label{fig:ghera}
    }
  \end{center}
\end{figure}
In Fig.~\ref{fig:ghera} we plot the shape of the gluon distribution $g(z) \equiv f_g(z)$
for representative kinematics of charm quark production at the HERA collider.
The typical values of $x$ and $Q$ employed for $xg[z(\eta)]$ in Fig.~\ref{fig:ghera} are correlated: 
The minimal (maximal) value of $x$ corresponds to the minimal (maximal) value of $Q$. 
For small scales $Q$ the value of $xg[z(\eta)]$ is suppressed at large $\eta$ due to its argument, 
because $z(\eta)$ rises with $\eta$ about linearly (see Eq.~(\ref{eq:F2c-eta})) 
and the gluon PDF decreases with rising argument.
Therefore, for small scales $Q$ the region of $\eta \lsim 1$ provides the
dominant contribution to the charm structure function $F_{2}^c$ 
and the parton kinematics relevant for the coefficient functions 
are effectively constrained to the threshold region (see Ref.~\cite{Vogt:1996wr}). 
Recall that $F_{2}^c$ is dominated by boson-gluon fusion.
On the other hand at large virtualities $Q$ the rise of $z(\eta)$ with $\eta$ is not so fast.
As a consequence, the suppression of the large-$\eta$ region due to the shape
of the gluon PDF is weaker, as can be seen in Fig.~\ref{fig:ghera}. 
However, for the case of bottom-quark production the large-$\eta$ suppression is stronger
due to the larger quark mass. 
Hence, even at large scales $Q$ the bottom structure function $F_{2}^b$  is still 
saturated by parton processes (and coefficient functions) close to threshold.

Let us now briefly summarize the threshold approximation to the hard
scattering coefficient functions in Eq.~(\ref{eq:coeff-exp}). 
To that end, we are following the standard procedure for resumming 
Sudakov logarithms, see e.g. Refs.~\cite{Contopanagos:1997nh,Kidonakis:1997gm}.
In differential kinematics for the one-particle inclusive DIS production
of a heavy quark (see~\cite{Laenen:1992zk,Laenen:1998kp}) the (dominant) gluon coefficient function is given by
\begin{eqnarray}
  \label{eq:ci0}
  c^{(i,0)}_{2,g\rm}(\eta, \xi) &=& 
  \int\limits_{s^{\prime}(1-\beta)/2}^{s^{\prime}(1+\beta)/2}\,d(-t_1)
  \int\limits_{0}^{s_4^{\rm max}}\,ds_4\, \quad
  K^{(i)}(s^{\prime}, t_1, u_1)\, 
  {d^2 c^{(0,0)}_{2,g}(s^{\prime}, t_1, u_1) \over dt_1\, ds_4} 
  \, ,
\end{eqnarray}
where $c^{(0,0)}_{2,g}$ is the Born contribution. We have $u_1=s^{\prime}+t_1-s_4$ and $s^{\prime}=s+\Q2$ and 
\begin{equation}
  \label{eq:s4max}
  s_4^{\rm max} = 
  {s \over s^{\prime}\, t_1}\, 
  \left(t_1 + {s^{\prime}(1-\beta) \over 2} \right)
  \left(t_1 + {s^{\prime}(1+\beta) \over 2} \right)
\, .
\end{equation}
In a physical interpretation $s_4$ denotes the additional energy carried away by soft gluon emission above the partonic threshold. 
At higher orders, Eq.~(\ref{eq:ci0}) contains plus-distributions of the type $\alpha_s^l\,[\ln^{2l-1}(s_4/\m2)/s_4]_+$ 
that give rise to the Sudakov logarithms upon integration, i.e. 
the well known double logarithms $\alphas^l \ln^{2l}\beta$ of an inclusive formulation.
At the differential level (one-particle inclusive kinematics) 
the threshold resummation for DIS heavy-quark production has been performed to NLL accuracy in Ref.~\cite{Laenen:1998kp}.
Subsequently, the resummed result has been used to generate the factors 
$K^{(i)}$ (cf. Eq.~(\ref{eq:ci0})) at fixed-order perturbation theory through NNLO.
These factors $K^{(i)}$ contain the large logarithms.

In the present paper we improve the approximate NNLO results of Ref.~\cite{Laenen:1998kp} by 
performing a matching at one-loop and by including the NLO Coulomb corrections.
This provides us the with the first three powers of Sudakov logarithms at all
orders and we arrive at the following expressions through NNLO,
\begin{eqnarray}
  \label{eq:K0}
  K^{(0)} &=& 
  \deltas4
  \, , \\
  \label{eq:K1}
  K^{(1)} &=& 
  {1 \over {4 \* \pi^2}} \* \biggl\{
         2 \* \ca \* \D1
       + \D0  \*  \biggl[
            \ca \* (\Lbeta + \ln(\t1/\u1) )
          - 2 \* \cf \* (1 + \Lbeta)
          \biggr]
       + \deltas4 \* \biggl[ 
          \cf \* \biggl(1 - \Lbeta 
  \\ 
  &&
          + 2 \* \ln(1 - \rs^2) \* \Lbeta + \ln(\rs) \* \Lbeta \biggr)
          + {1 \over 4} \* \ca \* \biggl(
          \ln(\t1/\u1)^2 - 3 \* \z2 - 4 \* \ln(1 - \rs^2) \* \Lbeta 
  \nonumber\\ 
  &&
          + 2 \* \ln(\rs) \* \Lbeta - 2 \* \ln(\rs) \* \ln(\t1/\u1) 
          - \ln(\rs)^2 
          + 2 \* \Li2(1- \u1/\t1/\rs) + 2 \* \Li2(1 - \t1/\u1/\rs)
          \biggr)
  \nonumber\\ 
  &&
       - \biggl(\cf-{\ca \over 2}\biggr)  \* 
         {(1-2 \* \m2/s) \over \beta}  \*  (\z2 - 2 \* \Li2(-\rs) - 2 \* \Li2(\rs))
         \biggr]
  \biggl\}
  \, , 
  \nonumber\\
  \label{eq:K2}
  K^{(2)} &=& 
  {1 \over {16 \* \pi^4}}\* \biggl\{
         2 \* \cas \* \D3
       + \D2  \*  \biggl[
            3 \* \cas \* (\Lbeta + \ln(\t1/\u1))
          - 6 \* \ca \* \cf \* (1 + \Lbeta)
          - {1 \over 2} \* \ca \* \b0
          \biggr]
  \\ 
  &&
       + \D1 \* \biggl[
            \ca \* K
          + 2 \* \ca \* \cf \* \biggl(
          1 - 3 \* \Lbeta - 2 \* \Lbeta^2 - 2 \* \ln(\t1/\u1) - 2 \* \ln(\t1/\u1) \* \Lbeta 
          + 2 \* \ln(1 - \rs^2) \* \Lbeta 
  \nonumber\\ 
  &&
          + \ln(\rs) \* \Lbeta
          \biggr)
          + \cf \* \b0 \* (1 + \Lbeta)
          + 4 \* \cfs \* (1 + \Lbeta)^2
          - {1 \over 2} \* \ca \* \b0 \* \biggl(
            \Lbeta + \ln(\t1/\u1)
          \biggr)
  \nonumber\\ 
  && 
          + {1 \over 2} \* \cas \* \biggl(
            2 \* \Lbeta^2 + 4 \* \ln(\t1/\u1) \* \Lbeta
            + 3 \* \ln(\t1/\u1)^2 + 4 \* \ln(-\u1/\m2) 
            - 11 \* \z2 
            - 4 \* \ln(1 - \rs^2) \* \Lbeta 
  \nonumber\\ 
  &&
            + 2 \* \ln(\rs) \* \Lbeta 
            - 2 \* \ln(\rs) \* \ln(\t1/\u1) 
            - \ln(\rs)^2 
            + 2 \* \Li2(1 - \u1/\t1/\rs) 
            + 2 \* \Li2(1 - \t1/\u1/\rs))
  \nonumber\\ 
  &&
       - 2 \* \biggl(\cf-{\ca \over 2}\biggr) \* \ca \* 
         {(1-2 \* \m2/s) \over \beta}  \* (\z2 - 2 \* \Li2(-\rs) - 2 \* \Li2(\rs))
         \biggr]
  \biggl\} 
  \, , 
  \nonumber
\end{eqnarray}
where $D_l = [\ln^{l}(s_4/\m2)/s_4]_+$ denote the plus-distribution. 
We have in QCD $\ca=3$, $\cf=4/3$, $\beta_0 = 11/3\*\ca-2/3\*\nf$ and 
$K= ( 67/18 - \z2 ) \ca - 5/9\nf$.
The variables and $\rs$ and $L_\beta$  are given by $\rs = (1-\beta)/(1+\beta)$ and 
$L_\beta = (1-2\,\m2/s)/\beta\, \left\{ \ln(\rs) + {\rm{i}}\pi \right\}$.

\begin{figure}[htb]
  \begin{center}
    \includegraphics[width=16.0cm]{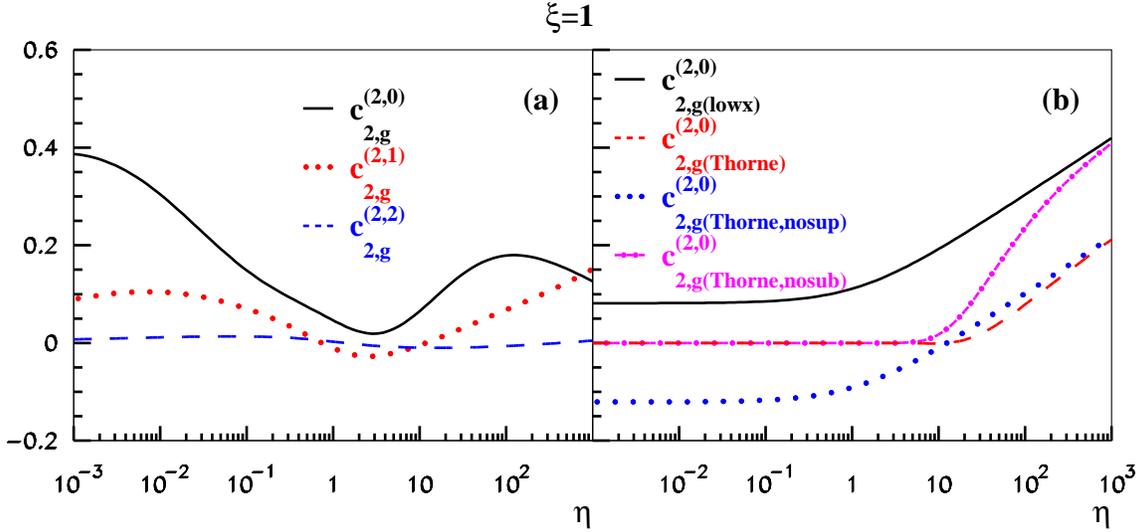}
    \vspace*{-8mm}
    \caption{ \small
      (a) The $\eta$-dependence of the gluon coefficient functions at NNLO 
      contributing to the heavy-quark DIS structure function $F_2$ (cf. Eq.~(\ref{eq:coeff-exp})).
      The solid line denotes $c^{(2,0)}_{2,g}$ according to Eq.~(\ref{eq:K2})
      dashes and dots show the exact result for $c^{(2,1)}_{2,g}$, 
      and $c^{(2,2)}_{2,g}$ from~\cite{Laenen:1998kp}.
      (b) The low-$x$ (large-$\eta$) asymptotics of $c_{2,g}^{(2,0)}$ 
      of Eq.~(\ref{eq:c2-smallx}) (solid line) from {\protect\cite{Catani:1990eg}} 
      and Eq.~(\ref{eq:c2-smallx-mod}) as modeled in {\protect\cite{Thorne:2006qt}} (dashes).
      Eq.~(\ref{eq:c2-smallx-mod}) without account of the model suppression factor (dashed dots) and  
      the model subtraction term (dots) are given for illustration.
      \label{fig:coef}}
  \end{center}
\end{figure}

Our improved NNLO approximation in Eq.~(\ref{eq:K2}) is exact in the region of phase space $s \simeq 4\m2$,
where perturbative corrections receive the largest weight from the convolution with
the gluon PDF (see discussion above and Refs.~\cite{Vogt:1996wr,Laenen:1998kp}).
In Eq.~(\ref{eq:ci0}) we have restricted ourselves to the case $\mus = \m2$. 
While it is straightforward to allow for general choices $\mus \neq \m2$ to
logarithmic accuracy in the threshold resummation formalism 
one can even derive the exact $\mur$ and $\muf$ scale dependence through NNLO~\cite{Laenen:1998kp} 
with the help of renormalization group methods. 
Thus, the functions $c^{(2,1)}_{2,g}$ and $c^{(2,2)}_{2,g}$ in Eq.~(\ref{eq:coeff-exp}) are known exactly~\cite{Laenen:1998kp},
see Fig.~\ref{fig:coef} for plots.

In a different kinematical regime, small-$x$ effects in DIS heavy-flavor
production have been studied systematically~\cite{Catani:1990eg} 
and incorporated in phenomenological analyses~\cite{Thorne:2006qt}.
Based on $k_{\rm T}$-resummation~\cite{Catani:1990eg}, the leading logarithm at small-$x$ for $c^{(2,0)}_{2,g}$ can be derived:
\begin{equation}
\label{eq:c2-smallx}
c^{(2,0)}_{2,g}(\eta, \xi) = {3 \over (2 \pi)^3}\, \ln(z/x)\, {\kappa_2(\xi) \over \xi}
\, ,
\end{equation}
recall $z(\eta)=x(1+4(\eta+1)/\xi)$.
The function $\kappa_2$  is a low order polynomial in $\xi=\Q2/\m2$ that 
can be determined from empirical fits to the leading $\ln(1/x)$ behavior of
Ref.~\cite{Catani:1990eg}. 
In the region of $\xi \lsim 1$ it has been estimated as~\cite{Thorne:2006qt}
\begin{equation}
\label{eq:kappa2}
\kappa_2(\xi) = 13.073 \* \xi - 23.827 \* \xi^2 + 24.107 \* \xi^3 - 9.173 \* \xi^4
\, .
\end{equation}
However, for phenomenological applications the sole knowledge of 
$\ln(x)$-terms is usually insufficient and additional assumption have to be supplied.
Thus, a particular functional form for the coefficient function at small-$x$
has been suggested in Ref.~\cite{Thorne:2006qt},
\begin{equation}
\label{eq:c2-smallx-mod}
c^{(2,0)}_{2,g}(\eta, \xi) = {3 \over (2 \pi)^3}\,  \beta\, \left( \ln(z/x) - 4 \right)\, (1-ax/z)^{20}\,{\kappa_2(\xi) \over \xi}
\end{equation}
where the subtraction term $(-4)$ at $\ln(1/z)$ is motivated by the terms of 
similar size in the small-$x$ limit of other known coefficient and splitting functions.
The factor $\beta$ times the polynomial $(1-ax/z)^{20}$ suppresses 
large-$z$ effects by a large power 
and $a$ is given below Eq.~(\ref{eq:totalF2c}). 

We plot the small-$x$ asymptotics of $c^{(2,0)}_{2,g}$ in Fig.~\ref{fig:coef}
where we display Eq.~(\ref{eq:c2-smallx}) for the original result of
Ref.~\cite{Catani:1990eg} and the model coefficient function of 
Eq.~(\ref{eq:c2-smallx-mod}). 
The variants of Eq.~(\ref{eq:c2-smallx-mod}) without account of the subtraction term $(-4)$ 
and the suppression factor of $\beta (1-ax/z)^{20}$ are also given for comparison. 
The form of Eq.~(\ref{eq:c2-smallx-mod}) is very sensitive to the particular choice of these terms. 
Due to the large-$z$ suppression Eq.~(\ref{eq:c2-smallx-mod}) vanishes at $\eta\lesssim 20$, 
and the region of $\eta$, where this happens is defined by the power of $20$ of the polynomial. 
Similarly, at large $\eta$ the form of Eq.~(\ref{eq:c2-smallx-mod}) is entirely 
defined by the subtraction term. 
The presence of subleading terms in the small-$x$ expansion with large numerical coefficients is a well documented feature 
at higher orders in QCD, see e.g. the example of the three-loop gluon splitting function~\cite{Vogt:2004mw}.
Moreover, precise phenomenological predictions based on small-$x$ approximations only
are extremely difficult to make, because the convolution in Eq.~(\ref{eq:F2c-eta}) is non-local. 
Thus, the small-$x$ terms in the coefficient function are weighted by the PDFs 
in the large-$x$ region (and vice versa), see~\cite{Vogt:2004mw}.

\begin{figure}[ht!]
  \begin{center}
    \includegraphics[width=14.0cm]{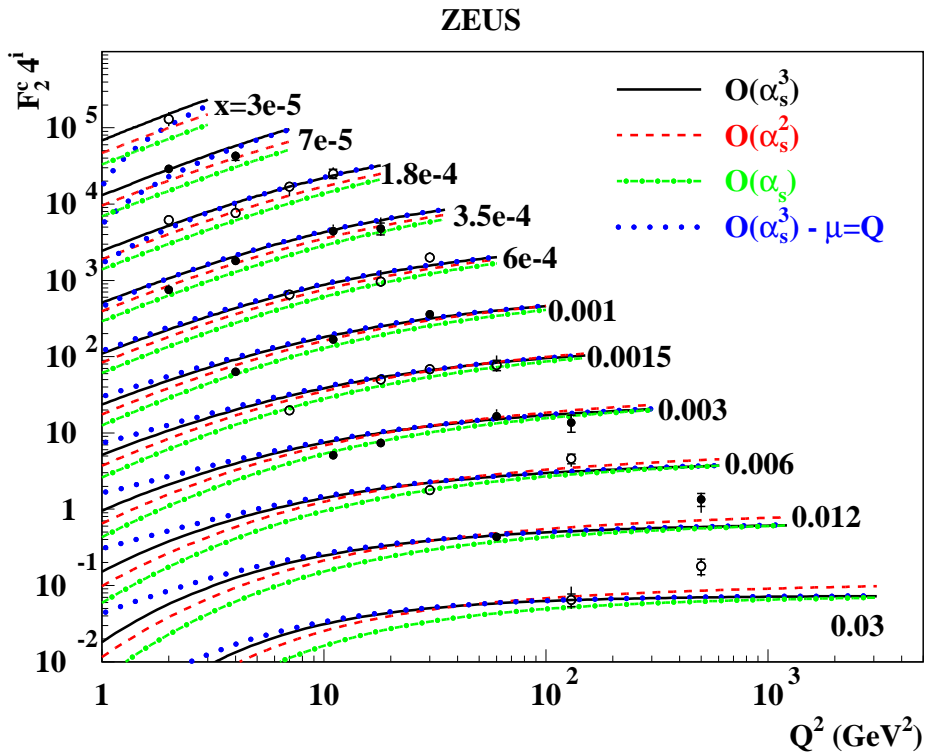}
    \vspace*{-8mm}
    \caption{ \small
      The predictions of the global fit of PDFs for the charm structure function $F_2^c$ 
      compared to data of Ref.~{\protect\cite{Chekanov:2003rb}} on $F_2^c$
      obtained in Run I of the HERA collider.
      The QCD corrections to the charm coefficient function in Eq.~(\ref{eq:coeff-exp}) 
      have been included up to NNLO (solid lines), NLO (dashes), and LO (dashed-dots) and the scale
      has been chosen $\mus=\Q2+4\m2$.
      The variant of the NNLO fit with the scale choice $\mu=Q$ is given by dots.
      \label{fig:zeus}
    }
  \end{center}
\end{figure}
For phenomenological applications our approximate NNLO result Eq.~(\ref{eq:K2}) for $c^{(2,0)}_{2,g}$ 
is added on top of the exact NLO predictions~\cite{Laenen:1992zk} and
supplemented with the exact NNLO scale dependent functions $c^{(2,1)}_{2,g}$
and $c^{(2,2)}_{2,g}$ of Ref.~\cite{Laenen:1998kp}. 
This provides the best present estimate for, say, the nucleon structure function $F_2^c$ 
for DIS electro-production of charm quarks.
We investigate the impact of these NNLO (gluon induced) contributions to $F_2^c$ 
on the nucleon PDFs extracted from global fits and perform a modified version 
of the fit of Ref.~\cite{Alekhin:2006zm}.
That fit is based on global data on inclusive charged-lepton DIS off nucleons 
supplemented by data for dimuon nucleon-nucleon production (i.e. the Drell-Yan process). 
Within the FFNS we take into account the NNLO corrections to the QCD evolution~\cite{Moch:2004pa,Vogt:2004mw} 
and the massless DIS and Drell-Yan coefficient functions~\cite{%
Kazakov:1992xj,vanNeerven:1991nn,Zijlstra:1991qc,Zijlstra:1992kj,Hamberg:1990np,Harlander:2002wh,Anastasiou:2003yy}. 
The value of $m$ was fixed at $1.25 \mbox{GeV}$, close to the world average and the factorization scale was selected as $\sqrt{Q^2+4m^2}$.
For comparision we also provide NNLO results for the factorization scale at $\mu=Q$. 
The latter choice naturally leads to larger deviations at small virtualities 
$Q$ due to a much increased numerical value of $\alpha_s$ (note that we have identified $\mu = \muf = \mur$).

\begin{figure}[ht!]
  \begin{center}
    \includegraphics[width=11.0cm]{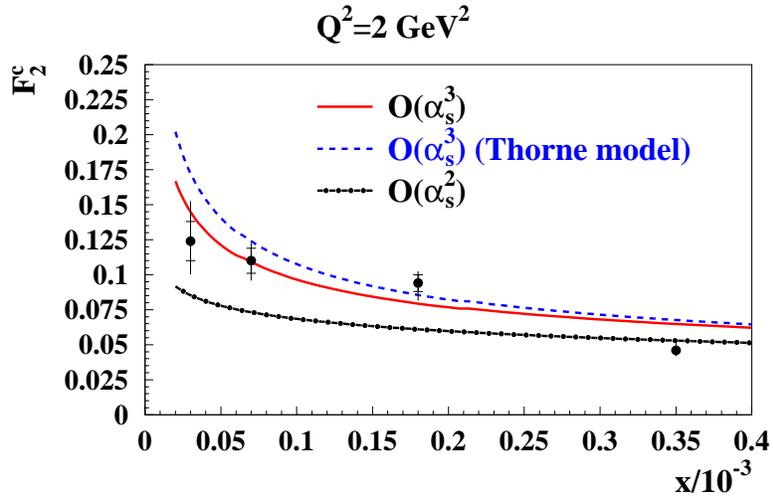}
    \vspace*{-8mm}
    \caption{ \small
      The impact of the large-$\eta$ tail of the NNLO coefficient
      functions on the value of $F_2^c$ at small $Q$ displaying our
      calculation (solid curve) and the same with 
      the model of Ref.~\cite{Thorne:2006qt} 
      for $c_{2,g}^{(2,0)}$ at large-$\eta$ added (dashes). 
      The NLO calculation (dashed-dots) and the data of
      Ref.\cite{Chekanov:2003rb} are given for comparison.
    \label{fig:higheta}
    }
  \end{center}
\end{figure}
We also perform two variants of this fit taking into account only the LO corrections of Refs.~\cite{Witten:1975bh,Gluck:1979aw} 
and the NLO corrections of Ref.~\cite{Laenen:1992zk}. 
The predictions for $F_2^c$ based on these three fits are compared in
Fig.~\ref{fig:zeus} to the ZEUS data of Ref.~\cite{Chekanov:2003rb}.  
The latter data are not used in the fits.
At the smallest values of $x$ and $Q$ in the plot the predictions rise
monotonically with increasing orders of perturbative QCD, thus improving agreement with the data. 
As we discussed above, in this region the value of $F_2^c$ is not sensitive 
to the coefficient functions at large $\eta$ and therefore our predictions
(cf. Eq.~(\ref{eq:K2})) can be considered as a good approximation to the full NNLO result for $F_2^c$.
At bigger values of $x$ and $Q$, as a result of a negative contribution from
$c_{2,g}^{(2,1)}$ at large $\eta$ the NNLO predictions dip below the NLO ones. 
In this region of $\eta$ the value of $c_{2,g}^{(2,0)}$ was set to zero as our
choice of matching the threshold approximation to fixed order perturbation theory 
(see e.g.~\cite{Moch:2008qy} for related discussions).
Checking the curves of Fig.~\ref{fig:zeus} at large values of $x$ and $Q$ 
one can conclude that this contribution should be positive in order to improve
the agreement with the data, the particular numerical impact
    depending, of course, on the gluon distribution shape as one can conclude from Fig.~\ref{fig:ghera}.
In this region (large $x$ and $Q$), the slope of $F_2^c$ appears to be distinctly flatter in $Q$, particularly at higher $x$. 
However, the kinematics for large values of $Q$ is far from threshold and
beyond control of our soft gluon approximation. 

The small-$x$ contribution to $c_{2,g}^{(2,0)}$ modeled in Ref.~\cite{Thorne:2006qt} 
affects the comparison to data at the lowest $Q$ and at small-$x$ only. 
In order to assess the impact of the small-$x$ term of
Eq.~(\ref{eq:c2-smallx}) quantitatively,  we focus on the lowest bin
$Q^2=2~{\mbox{GeV}}$ and illustrate its effect in Fig.~\ref{fig:higheta}.
At the lowest value in $x$ we do observe a slight sensitivity 
on the small-$x$ term, which in terms of $\eta(z)$ corresponds to the region of larger $\eta \gg1$. 
The effect of the high-$\eta$ model of Ref.~\cite{Thorne:2006qt} amounts
at most to a $30 \%$-fraction of the dominant NNLO contribution coming from the
threshold region at the lowest values of $x$. 
However, in this region, the addition of the small-$x$ contribution to
$c_{2,g}^{(2,0)}$ overshoots the data while it vanishes quickly at larger $x$.
Recall in our analysis we set the contribution of the threshold logarithms in $c_{2,g}^{(2,0)}$ to zero for $\eta > 1$.
As we discussed above, the ansatz of Eq.~(\ref{eq:c2-smallx}) 
has an inherent model uncertainty of $100\%$ since it is driven by ad hoc parameters. 
Therefore it cannot be used in quantitative comparisons. 

\begin{figure}[hbt]
  \begin{center}
    \includegraphics[width=14.0cm]{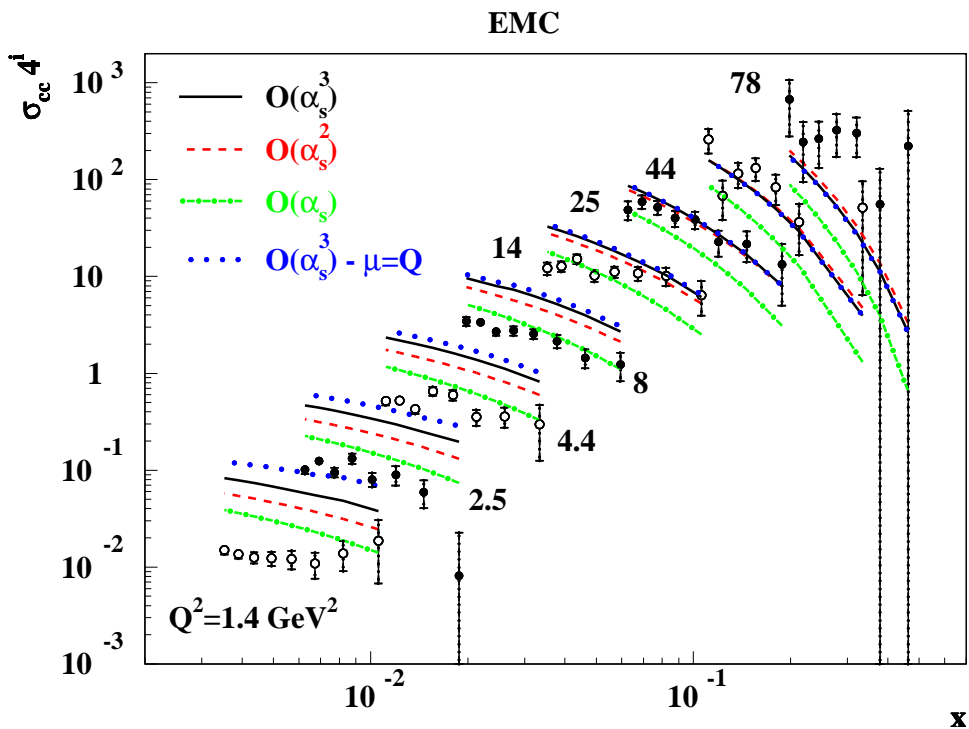}
    \vspace*{-8mm}
    \caption{ \small
      The same as Fig.~\ref{fig:zeus}
      for the charm electro-production 
      cross section data obtained by the EMC experiment
      {\protect\cite{Aubert:1982tt}}.
      \label{fig:emc}
    }
  \end{center}
\end{figure}
At fixed-target energies the charm contribution to the inclusive
sample is much smaller than at the HERA collider. 
This makes the experimental determination of $F_2^c$ more difficult. 
The only conclusive data on fixed-target charm electro-production were
obtained some time ago by the EMC collaboration~\cite{Aubert:1982tt}. 
These data are compared to the predictions of our fits in Fig.~\ref{fig:emc}. 
At smallest values of $Q$ the data are in disagreement with the predictions. 
Moreover the disagreement increases with the order of the perturbative QCD
correction. 
At bigger values of $Q$ the difference between the NLO and the NNLO predictions 
is marginal due to the difference in the gluon PDFs obtained in these variants of the fit. 
However the general agreement between data and calculations is far from ideal. 
Due to the limited collision energy the EMC data are sensitive to the region of $\eta<1$ only, 
where our NNLO approximation in Eq.~(\ref{eq:K2}) should describe the exact
coefficient function $c_{2,g}^{(2,0)}$ very well.
Thus, we see no way to improve the agreement with the EMC data by performing
a complete calculation of $c_{2,g}^{(2,0)}$.
Since the EMC data are unique it seems to be useful to have additional experimental 
input to clarify this disagreement.

\begin{figure}[htb]
  \begin{center}
    \includegraphics[width=16.0cm]{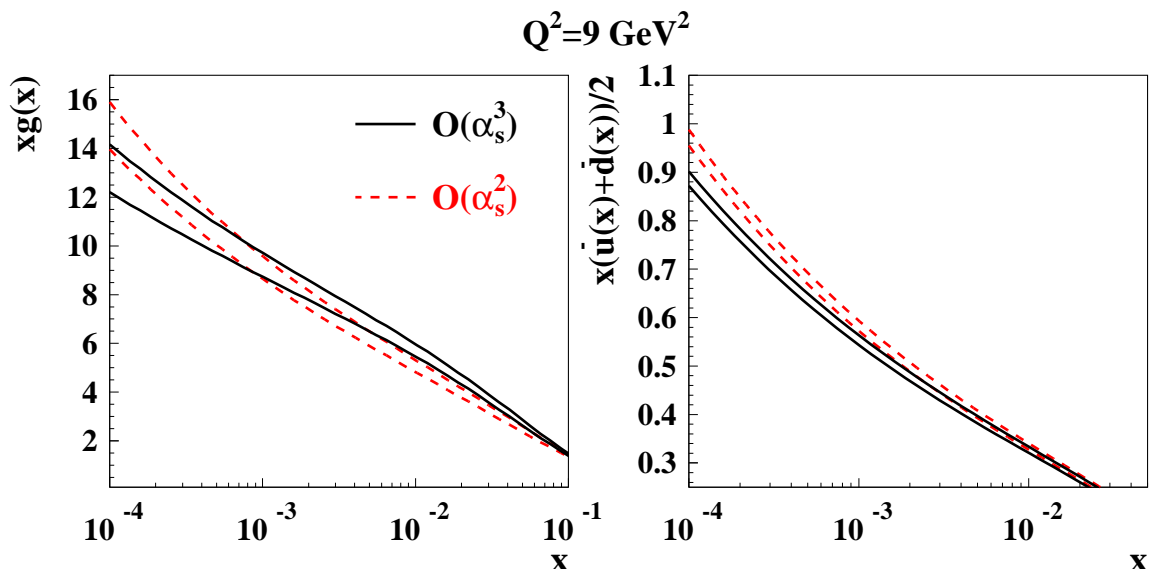}
    \vspace*{-5mm}
    \caption{ \small
      The $\pm 1\sigma$ band for the gluon (left panel) and non-strange sea
      (right panel) distributions obtained in the fit by including 
      the coefficient functions for charm electro-production up to NNLO 
      (solid lines) and NLO (dashes).
      \label{fig:pdfs}
    }
  \end{center}
\end{figure}
Despite the fact that we do not use data for the charm electro-production
in the fit our results are sensitive to the details of the description of $F_2^c$.
This is because $F_2^c$ amounts to a substantial contribution, up to $30\%$, 
to the inclusive DIS structure functions at the HERA collider energies. 
The biggest variation in the fitted PDFs due to the NNLO corrections to $F_2^c$ 
is observed for the sea quarks PDF at small-$x$. 
The latter becomes smaller in order to compensate the positive contribution 
from the NNLO term at small-$x$ (see Fig.~\ref{fig:pdfs}). 
The gluon distribution at $x\sim 0.02$ becomes bigger by about one standard deviation 
as a compensation of the negative contribution of the NNLO term 
and the fitted value of $\alpha_{\rm s}(M_{\rm Z})$, 
which is anti-correlated with the gluon distribution at small-$x$, 
goes down by about $1\sigma$. 
Other PDFs are essentially not affected by the corrections.

Let us summarize: 
We have improved the perturbative QCD predictions for the heavy-quark DIS structure functions.
Our NNLO approximation takes along the first three powers of Sudakov logarithms 
for the boson-gluon channel $\gamma g$, which, in an inclusive formulation (performing the integration and keeping the
leading terms in $\beta$ only) corresponds to all logarithmically enhanced terms $\ln^k\beta$, $k=2,\dots,4$.
Moreover, we have employed the exact expressions for all scale dependent terms through NNLO~\cite{Laenen:1998kp}.

Subsequently, we have applied the NLO QCD corrections~\cite{Laenen:1992zk} to
the charm structure function $F_2^c$ and our new approximate NNLO result in a global fit
to data for charged-lepton DIS and dimuon production in the Drell-Yan process.
Especially the use of the threshold-approximated NNLO result is legitimate 
because the gluon PDF constrains the parton kinematics to values around $s \simeq 4 \m2$.
Modifying the fit of Ref.~\cite{Alekhin:2006zm} in this way we have studied the effects 
for the determination of PDFs.
We have found that our approximate NNLO result for $F_2^c$  gives better agreement between 
the fitted PDFs and the HERA collider data. 
The agreement with data extends even down to small values of $x$.
The results of our fit are also in good agreement with ZEUS data~\cite{Chekanov:2003rb} 
on charm electro-production, which was not used in the fits.
Comparing with EMC data~\cite{Aubert:1982tt} we did find disagreement, though, 
and it would be interesting to get new and independent experimental
information in order to resolve it.

On the theory side we could in principle extend the NNLO threshold approximation 
of Eq.~(\ref{eq:K2}) further to include (after integration) the linear logarithm in $\ln\beta$ at
two loops and the two-loop Coulomb corrections following
the procedure of Ref.~\cite{Moch:2008qy} for heavy-quark hadro-production.
However, we leave this to future research.

\subsection*{Acknowledgments}
We would like to thank R.~Thorne for communication on Ref.~\cite{Thorne:2006qt}.
S.A. is supported by the RFBR grant 02-06-16659 and 
S.M. by the Helmholtz Gemeinschaft under contract VH-NG-105.
This work is also partly supported by DFG in SFB/TR 9.



{\footnotesize
\begin{thebibliography}{10}

\bibitem{Witten:1975bh}
E. Witten,
\newblock Nucl. Phys. B104 (1976) 445
\newblock 

\bibitem{Gluck:1979aw}
M. Gl\"uck and E. Reya,
\newblock Phys. Lett. B83 (1979) 98
\newblock 

\bibitem{Aktas:2005iw}
H1, A.~Aktas et al.,
\newblock Eur. Phys. J.\  C45 (2006) 23, hep-ex/0507081
\newblock 

\bibitem{Chekanov:2003rb}
ZEUS, S. Chekanov et~al.,
\newblock Phys. Rev. D69 (2004) 012004, hep-ex/0308068
\newblock 

\bibitem{Alekhin:2002fv}
S. Alekhin,
\newblock Phys. Rev. D68 (2003) 014002, hep-ph/0211096;
\newblock 
\newblock JETP Lett. 82 (2005) 628, hep-ph/0508248
\newblock 

\bibitem{Alekhin:2006zm}
S. Alekhin, K. Melnikov and F. Petriello,
\newblock Phys. Rev. D74 (2006) 054033, hep-ph/0606237
\newblock 

\bibitem{Thorne:2008xf}
R.S. Thorne and W.K. Tung,
\newblock (2008), 0809.0714
\newblock 

\bibitem{Laenen:1992zk}
E. Laenen et~al.,
\newblock Nucl. Phys. B392 (1993) 162
\newblock 

\bibitem{Kazakov:1992xj}
D.I. Kazakov and A.V. Kotikov,
\newblock Phys. Lett. B291 (1992) 171
\newblock 

\bibitem{vanNeerven:1991nn}
W.L. van Neerven and E.B. Zijlstra,
\newblock Phys. Lett. B272 (1991) 127
\newblock 

\bibitem{Zijlstra:1991qc}
E.B. Zijlstra and W.L. van Neerven,
\newblock Phys. Lett. B273 (1991) 476
\newblock 

\bibitem{Zijlstra:1992kj}
E.B. Zijlstra and W.L. van Neerven,
\newblock Phys. Lett. B297 (1992) 377
\newblock 

\bibitem{Moch:1999eb}
S. Moch and J.A.M. Vermaseren,
\newblock Nucl. Phys. B573 (2000) 853, hep-ph/9912355
\newblock 

\bibitem{Moch:2004pa}
S. Moch, J.A.M. Vermaseren and A. Vogt,
\newblock Nucl. Phys. B688 (2004) 101, hep-ph/0403192
\newblock 

\bibitem{Vogt:2004mw}
A. Vogt, S. Moch and J.A.M. Vermaseren,
\newblock Nucl. Phys. B691 (2004) 129, hep-ph/0404111
\newblock 

\bibitem{Shifman:1977yb}
M.~A.~Shifman, A.~I.~Vainshtein and V.~I.~Zakharov,
\newblock Nucl. Phys. B136 (1978) 157; Yad. Fiz. 27 (1978) 455
\newblock 

\bibitem{Buza:1995ie}
M. Buza et~al.,
\newblock Nucl. Phys. B472 (1996) 611, hep-ph/9601302
\newblock 

\bibitem{Bierenbaum:2007qe}
I. Bierenbaum, J. Bl\"umlein and S. Klein,
\newblock Nucl. Phys. B780 (2007) 40, hep-ph/0703285
\newblock 

\bibitem{Chuvakin:1999nx}
A. Chuvakin, J. Smith and W.L. van Neerven,
\newblock Phys. Rev. D61 (2000) 096004, hep-ph/9910250
\newblock 

\bibitem{Tung:2001mv}
W.K. Tung, S. Kretzer and C. Schmidt,
\newblock J. Phys. G28 (2002) 983, hep-ph/0110247
\newblock 

\bibitem{Gluck:1993dpa}
M.~Gl\"uck, E.~Reya and M.~Stratmann,
\newblock Nucl. Phys. B422 (1994) 37
\newblock 

\bibitem{Laenen:1998kp}
E. Laenen and S. Moch,
\newblock Phys. Rev. D59 (1999) 034027, hep-ph/9809550
\newblock 

\bibitem{Vogt:1996wr}
A. Vogt,
\newblock (1996), hep-ph/9601352
\newblock 

\bibitem{Contopanagos:1997nh}
H. Contopanagos, E. Laenen and G. Sterman,
\newblock Nucl. Phys. B484 (1997) 303, hep-ph/9604313
\newblock 

\bibitem{Kidonakis:1997gm}
N. Kidonakis and G. Sterman,
\newblock Nucl. Phys. B505 (1997) 321, hep-ph/9705234
\newblock 

\bibitem{Catani:1990eg}
S. Catani, M. Ciafaloni and F. Hautmann,
\newblock Nucl. Phys. B366 (1991) 135
\newblock 

\bibitem{Thorne:2006qt}
R.S. Thorne,
\newblock Phys. Rev. D73 (2006) 054019, hep-ph/0601245
\newblock 

\bibitem{Hamberg:1990np}
R. Hamberg, W.L. van Neerven and T. Matsuura,
\newblock Nucl. Phys. B359 (1991) 343
\newblock 

\bibitem{Harlander:2002wh}
R.V. Harlander and W.B. Kilgore,
\newblock Phys. Rev. Lett. 88 (2002) 201801, hep-ph/0201206
\newblock 

\bibitem{Anastasiou:2003yy}
C. Anastasiou et~al.,
\newblock Phys. Rev. Lett. 91 (2003) 182002, hep-ph/0306192
\newblock 

\bibitem{Moch:2008qy}
S. Moch and P. Uwer,
\newblock Phys. Rev. D78 (2008) 034003, 0804.1476
\newblock 

\bibitem{Aubert:1982tt}
EMC, J.J. Aubert et~al.,
\newblock Nucl. Phys. B213 (1983) 31
\newblock 

\end{thebibliography}

}
\end{document}